\documentclass[reprint,aps]{revtex4-1}
\usepackage{amsmath}
\usepackage{graphicx}
\usepackage{dcolumn}
\usepackage{bm}

\begin{document}

\title{Pinned modes in lossy lattices with local gain and nonlinearity}
\author{Boris A. Malomed}
\affiliation{Department of Physical Electronics, School of
Electrical Engineering, Faculty of Engineering, Tel Aviv University,
Tel Aviv 69978, Israel}
\author{Edwin Ding}
\affiliation{Department of Mathematics and Physics, Azusa Pacific
University, Box 7000, Azusa, CA 91702-7000, USA}
\author{K. W. Chow and S. K. Lai}
\affiliation{Department of Mechanical Engineering, University of
Hong Kong, Pokfulam Road, Hong Kong}

\begin{abstract}
We introduce a discrete linear lossy system with an embedded ``hot
spot" (HS), i.e., a site carrying linear gain and complex cubic
nonlinearity. The system can be used to model an array of optical or
plasmonic waveguides, where selective excitation of particular cores
is possible. Localized modes pinned to the HS are constructed in an
implicit analytical form, and their stability is investigated
numerically. Stability regions for the modes are obtained in the
parameter space of the linear gain and cubic gain/loss. An essential
result is that the interaction of the \emph{unsaturated} cubic gain
and self-defocusing nonlinearity can produce stable modes, although
they may be destabilized by finite-amplitude perturbations. On the
other hand, the interplay of the cubic loss and self-defocusing
gives rise to a bistability.
\end{abstract}

\maketitle

\section{Introduction}

Dissipative spatial solitons, which originate from the interaction of
diffraction, self-focusing nonlinearity, and dissipation (gain and loss),
have drawn great interest in nonlinear optics~\cite{Rosanov} and, more
recently, in plasmonics~\cite{plasmonics,Marini}. A necessary condition
required for the formation of stable solitons is the stability of the zero
background. The simplest single-component complex Ginzburg-Landau (CGL)
equation with the uniform linear gain is not an appropriate candidate for
modeling dissipative solitons, as it violates this condition. Dissipative
solitons may be stabilized by a system of linearly coupled CGL equations~%
\cite{wg1} that models dual-core waveguides with the linear gain and loss
acting in different cores~\cite{dual,Pavel,Chaos,Marini}. Stable solitons
can also be generated by the single CGL equations that incorporate the
linear loss, cubic gain and quintic loss, which accounts for the nonlinear
saturable absorption~\cite{CQ,cqgle}. In these models, the quintic loss
saturates the growth induced by the cubic gain and therefore stabilizes the
solitons.

Another method for generating stable localized modes, which has
recently drawn considerable attention, relies on the action of
linear gain at a ``hot spot" (HS, i.e., a localized region in a
lossy
waveguide~\cite{HSexact,HS,Valery} or in a dissipative Bragg grating~\cite%
{Mak}). Models with several hot spots~\cite{spotsExact1,spots,spotsExact2},
as well as with similar extended structures~\cite{Zezyu}, have also been
studied. HSs can be created by implanting an appropriate distribution of
gain-producing dopants into the waveguide~\cite{Kip}, or by illuminating a
uniformly-doped waveguide with external pump beam(s) focused at the
designated spot(s). Dissipative solitons pinned to HSs can be stabilized via
the balance between the local amplification and uniform loss in the
waveguide. In particular, solutions for dissipative solitons pinned to
narrow HSs approximated by delta-functions have been found analytically~\cite%
{HSexact,spotsExact1,spotsExact2}. Other relevant modes, both one- and
two-dimensional, including stable vortices supported by the gain applied to
a confined area~\cite{2D}, have been generated in the numerical form~\cite%
{HS,Valery,spots}. Stable dissipative solitons have also been predicted in a
system that combines the uniformly-distributed linear gain and nonlinear
loss growing towards the periphery faster than $r^{D}$, where $r$ and $D$
represent the distance from the center and spatial dimension, respectively~%
\cite{Barcelona}.

An interesting ramification of the configurations mentioned above is the
possibility to generate stable solitons supported by localized cubic gain,
in the absence of the quintic gain saturation. While dissipative solitons
cannot be stable without higher-order nonlinear losses in uniform media~\cite%
{Kramer,ml}, it was recently demonstrated~\cite{Valery} that \emph{stable}
dissipative solitons may be pinned to an HS carrying the unsaturated cubic
gain and embedded into a uniform linear-loss background. This finding
suggests ways to design clean nonlinear soliton amplification that avoids
concomitant generation of noise, which is also relevant for plasmonics~\cite%
{Korea}.

The present work explores the generation of stable solitons in discrete
waveguiding arrays (lattices) with a localized unsaturated nonlinear gain.
In particular, we demonstrate that this is possible in a linear lattice
where the nonlinearity, represented by the self-phase modulation\ and cubic
gain, is applied to a single waveguide (the HS). The lattice CGL system is
introduced here as a discrete counterpart of the continuous HS models~\cite%
{HSexact,Valery}, and can be used to investigate various effects in
photonics, cf. Refs.~\cite{review, discrCGL,Eyal}. In particular, it may be
used for selective excitation of particular core(s) in an arrayed
waveguiding system, if it is uniformly doped, but only the selected core is
pumped by an external coherent source of light. In addition to allowing the
straightforward experimental implementation~\cite{review}, we demonstrate
that the present discrete system supports analytical solutions for discrete
solitons (similar to those in the discrete linear Schr\"{o}dinger equation
with embedded nonlinear elements~\cite{embed}), thus providing a natural
platform for exploring the soliton dynamics.

The paper is organized as follows. The discrete CGL equation with HS, and
its implicit analytical solutions for pinned modes, are introduced in Sec.~%
\ref{sec:gov}. The linear stability analysis of the solitons against small
perturbations is presented in Sec.~\ref{sec:stab}, and results of numerical
computations of the corresponding eigenvalue spectra are reported in Sec.~%
\ref{sec:num}. The predictions of the linear-stability analysis are compared
to direct simulations of perturbed evolution of the discrete solitons. In
particular, stable solitons are found under the \emph{unsaturated nonlinear
local gain}, provided that the localized nonlinearity is self-defocusing,
although finite-amplitude perturbations may destabilize these modes. On the
other hand, the interplay of the self-defocusing nonlinearity and cubic loss
gives rise to bistability of the pinned modes, which is a sufficiently
interesting finding too. The paper is concluded by~Sec. \ref{sec:con}.

\section{The model}

\label{sec:gov}

The present work is motivated by the complex Ginzburg-Landau (CGL) equation
that models the propagation of an electromagnetic field of amplitude $u(x,z)$
in a lossy waveguide with an embedded HS:
\begin{equation}
\frac{\partial u}{\partial z}=\frac{i}{2}\frac{\partial ^{2}u}{\partial x^{2}%
}-\gamma u+\left[ \left( \Gamma _{1}+i\Gamma _{2}\right) +\left( iB-E\right)
|u|^{2}\right] \delta (x)u\;.  \label{eq:cgl}
\end{equation}%
Here $x$ and $z$ are the transverse and longitudinal coordinates, and $%
\gamma >0$ is the background linear loss. Delta-function $\delta (x)$
approximates the concentration of the linear gain ($\Gamma _{1}>0$), linear
potential ($\Gamma _{2}>0$ corresponds to the local attraction), cubic
dissipation $E$ (positive and negative for the loss and gain, respectively),
and Kerr nonlinearity $B$ (positive and negative for the self-focusing and
defocusing) at the HS.

As the model of an array of guiding cores, replacing a single continual
waveguide, Eq.~(\ref{eq:cgl}) is substituted by its discrete version,
\begin{gather}
\frac{\mathrm{d}u_{m}}{\mathrm{d}z}=\frac{i}{2}\left(
u_{m-1}-2u_{m}+u_{m+1}\right) -\gamma u_{m}  \notag \\
+\left[ \left( \Gamma _{1}+i\Gamma _{2}\right) +\left( iB-E\right)
|u_{m}|^{2}\right] \delta _{m,0}u_{m}\;,  \label{eq:gl}
\end{gather}%
where $m=0$, $\pm 1$, $\pm 2$, ... is the discrete coordinate, $\delta _{m,0}
$ is the Kronecker's symbol, and the coefficient of the linear coupling
between adjacent cores is scaled to unity. In optics, the discrete equation
can be derived \ by means of well-known methods \cite{Chr,discrCGL,review}.
In the application to arrays of plasmonic waveguides, which can be built,
for example, as a set of metallic nanowires mounted on top of a dielectric
structure \cite{plasmon-array}, this equation can be derived in the
adiabatic approximation, when the exciton field may be eliminated in favor
of the photonic component (otherwise, the discrete system will be
two-component). It is also relevant to mention that the well-known \textit{%
staggering transformation} \cite{review}, $u_{m}(t)\equiv \left( -1\right)
^{m}e^{-2it}\tilde{u}_{m}^{\ast }$, where the asterisk stands for the
complex conjugate, simultaneously reverses the signs of $\Gamma _{2}$ and $B$%
, thus rendering the self-focusing and defocusing nonlinearities mutually
convertible in the discrete system. In particular, the latter feature is
essential for modeling arrays of plasmonic waveguides, where the intrinsic
excitonic nonlinearity is self-repulsive. In what follows, we fix the signs
of $\Gamma _{2}$ and $B$ by defining $\Gamma _{2}>0$, while $B$ may be
positive (self-focusing), negative (self-defocusing), or zero.

As mentioned above, the underlying array can be actually manufactured as a
uniform one, with all the cores doped by an appropriate amplifying material,
while the HS is singled out by focusing an external pump to a single core.
The latter setting is interesting for potential applications, as the
location of the HS is switchable.

The model based on Eq. (\ref{eq:gl}) is the subject of the present paper.
The Kerr-nonlinearity coefficient, if present, may be normalized to $B=+1$
(self-focusing) or $B=-1$ (self-defocusing). These two cases are considered
separately in Sec.~\ref{sec:num}, along with the case of $B=0$, when the
nonlinearity is represented solely by the cubic dissipation localized at the
HS.

Dissipative solitons in uniform discrete CGL equations were studied by means
of numerical methods in Refs. \cite{discrCGL,Eyal}. We seek analytical
solutions for stationary modes with real propagation constant $k$ as
\begin{equation}
u_{m}=U_{m}e^{ikz}.  \label{eq:pw}
\end{equation}%
Outside of the HS site, $m=0$, the linear lattice gives rise to the exact
solution with real amplitude $A$,
\begin{equation}
U_{m}=A\exp (-\lambda |m|),\;|m|~\geq 1,  \label{eq:exp}
\end{equation}%
and complex $\lambda \equiv \lambda _{1}+i\lambda _{2}$, localized modes
corresponding to $\lambda _{1}>0$.

The amplitude at the HS (central) site may be different from $A$, and hence
we assume
\begin{equation}
U_{0}=A_{R}+iA_{I},  \label{eq:U0}
\end{equation}%
for some real constants $A_{R}$ and $A_{I}$. Substituting~Eqs. (\ref{eq:pw}%
), (\ref{eq:exp}) and~(\ref{eq:U0}) into the discrete CGL~(\ref{eq:gl})
yields six nonlinear algebraic equations for $A$, $A_{R}$, $A_{I}$, $\lambda
_{1}$, $\lambda _{2}$ and $k$. Straightforward considerations demonstrate
that any solution has $A_{R}=A$ and $A_{I}=0$, hence the six equations
reduce to four:
\begin{eqnarray}
-1+\cosh \lambda _{1}\cos \lambda _{2} &=&k,  \notag \\
-\gamma -\sinh \lambda _{1}\sin \lambda _{2} &=&0,  \notag \\
e^{-\lambda _{1}}\sin \lambda _{2}-\gamma +\Gamma _{1}-EA^{2} &=&0,  \notag
\\
e^{-\lambda _{1}}\cos \lambda _{2}-1+\Gamma _{2}+BA^{2} &=&k.  \label{eq:sys}
\end{eqnarray}

\begin{figure}[t]
\begin{center}
\includegraphics[width = 80mm, keepaspectratio]{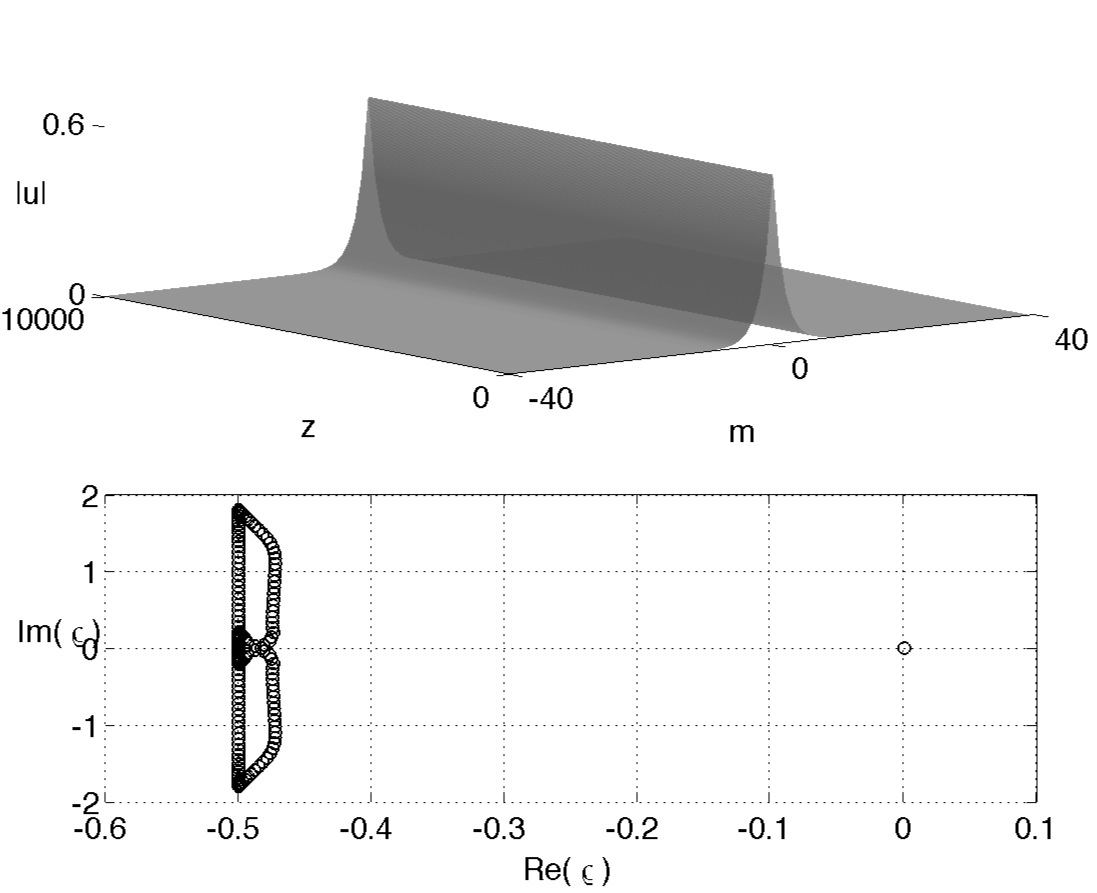}
\end{center}
\caption{Top: The evolution of the pinned-soliton solution~(\protect\ref%
{eq:pw}) and~(\protect\ref{eq:exp}) with $A=0.597$, $\protect\lambda %
=0.626-0.846i$, and $k=-0.202$. The system's parameters are $\protect\gamma %
=0.5$, $\Gamma _{1}=0.9$, $\Gamma _{2}=0.8$, $B=-1$, and $E=0$. Bottom: The
spectrum of stability eigenvalues for small perturbations around this
solution. All the eigenvalues have non-positive real parts, hence the pinned
mode is stable.}
\label{fig:ex}
\end{figure}

This system of algebraic equations can be solved numerically by means of the
Newton's method for $A$, $\lambda _{1}$, $\lambda _{2}$, and $k$. For
instance, with $\gamma =0.5$, $\Gamma _{1}=0.9$, $\Gamma _{2}=0.8$, $B=-1$,
and $E=0$ (these parameters correspond to the self-defocusing local
nonlinearity and zero cubic loss), a physically relevant solution is $A=0.597
$, $\lambda _{1}=0.63$, $\lambda _{2}=-0.85$, and $k=-0.2$. The top panel of
Fig.~\ref{fig:ex} shows the stable evolution of the corresponding mode,
produced by simulating Eq.~(\ref{eq:gl}) with the help of the fourth-order
Runge-Kutta algorithm, using periodic boundary conditions.

Families of pinned modes~(\ref{eq:pw}) and their stability are presented in
detail below. Linear gain $\Gamma _{1}$ and cubic gain/loss $E$ are used as
control parameters in the analysis (in the physical system outlined above,
their values may be adjusted by varying the intensity of the external pump).
Note that, in the case of $B=E=0$, amplitude $A$ becomes arbitrary and drops
out from Eqs.~(\ref{eq:sys}), as Eq.~(\ref{eq:gl}) becomes linear. In this
case, $\Gamma _{1}$ may be considered as another unknown, determined by the
balance between the background loss and localized gain in the linear system,
which implies the structural instability of the stationary trapped modes in
the linear model. In the presence of the nonlinearity, the power balance can
be adjusted through the value of the amplitude at given $\Gamma _{1}$,
therefore solutions, including stable ones, can be found in a range of
values of the linear gain, $\Gamma _{1}$.

\section{The linear-stability analysis}

\label{sec:stab}

The stability of the pinned modes was studied by means of the linearization
procedure~\cite{ted}. To this end, perturbed solutions were taken as
\begin{equation}
u_{m}=\left[ U_{m}+\epsilon V_{m}(z)\right] e^{ikz}\;,  \label{eq:ansatz}
\end{equation}%
where $V_{m}(z)=X_{m}(z)+iY_{m}(z)$ is a complex perturbation with an
infinitesimal amplitude $\epsilon \ll 1$. Substituting this into Eq.~(\ref%
{eq:gl}), one derives the following linear equations:%
\begin{eqnarray}
&&\frac{\mathrm{d}X_{m}}{\mathrm{d}z}=-\frac{1}{2}Y_{m-1}+(k+1)Y_{m}-\frac{1%
}{2}Y_{m+1}-\gamma X_{m}  \notag \\
&&+\delta _{m,0}\left\{ \left( \Gamma _{1}X_{m}-\Gamma _{2}Y_{m}\right)
\right.   \notag \\
&&-B\left[ 2P_{m}Q_{m}X_{m}+\left( P_{m}^{2}+3Q_{m}^{2}\right) Y_{m}\right]
\notag \\
&&\left. -E\left[ \left( 3P_{m}^{2}+Q_{m}^{2}\right) X_{m}+2P_{m}Q_{m}Y_{m}%
\right] \right\} \;,  \notag \\
&&\frac{\mathrm{d}Y_{m}}{\mathrm{d}z}=\frac{1}{2}X_{m-1}-(k+1)X_{m}+\frac{1}{%
2}X_{m+1}-\gamma Y_{m}  \notag \\
&&+\delta _{m,0}\left\{ \left( \Gamma _{2}X_{m}+\Gamma _{1}Y_{m}\right)
\right.   \notag \\
&&+B\left[ \left( 3P_{m}^{2}+Q_{m}^{2}\right) X_{m}+2P_{m}Q_{m}Y_{m}\right]
\notag \\
&&\left. -E\left[ 2P_{m}Q_{m}X_{m}+\left( P_{m}^{2}+3Q_{m}^{2}\right) Y_{m}%
\right] \right\} \;,  \label{eq:linear}
\end{eqnarray}%
where $P_{m}\equiv \mathrm{Re}(U_{m})$ and $Q_{m}\equiv \mathrm{Im}(U_{m})$.
An eigenvalue problem is obtained by substituting $X_{m}=\phi _{m}\exp (\rho
z)$ and $Y_{m}=\psi _{m}\exp (\rho z)$ into Eqs.~(\ref{eq:linear}). The
pinned mode is linearly stable provided that all the eigenvalues have $%
\mathrm{Re}\left( \rho \right) \leq 0$. An example of the stable numerically
calculated spectrum for the stationary mode considered in the previous
subsection is shown in the bottom panel of Fig.~\ref{fig:ex}.

\section{Numerical results}

\label{sec:num}

With the help of the methods outlined above, we consider three different
cases: (i) the self-defocusing nonlinearity ($B=-1$), (ii) the self-focusing
nonlinearity ($B=+1$), and (iii) zero nonlinearity ($B=0$). In each case,
the cubic gain ($E<0$) and loss ($E>0$) are investigated separately.

\subsection{The self-defocusing regime ($B=-1$)}

\begin{figure}[t]
\begin{center}
\includegraphics[width = 80mm, keepaspectratio]{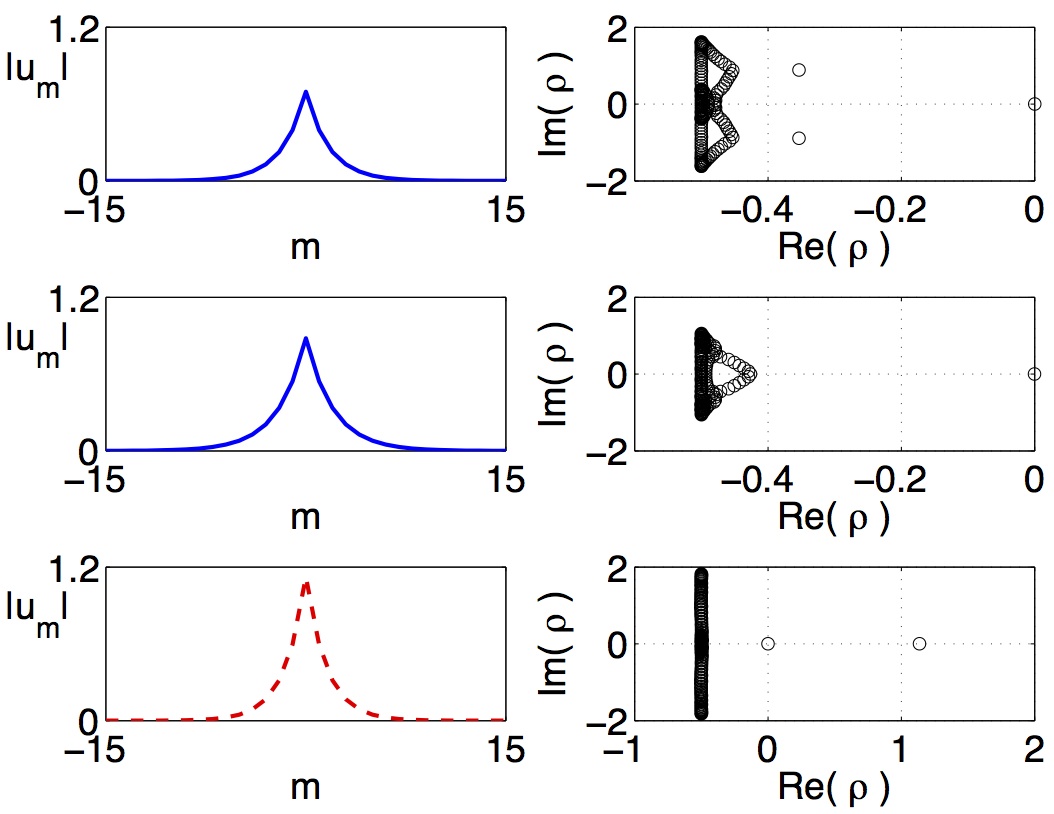}
\end{center}
\caption{(Color online) Examples of stable and unstable pinned solitons
(left) and the corresponding stability spectra (right) in the case of the
self-defocusing nonlinearity combined with the cubic unsaturated gain: $B=-1$%
, $E=-0.016$, $\protect\gamma =0.5$, and $\Gamma _{2}=0.8$. Top: A stable
solution found at $\Gamma _{1}=0.9048$. Middle: The stable solution with the
largest amplitude, found at $\Gamma _{1}=0.9936$. Bottom: An unstable
solution obtained at $\Gamma _{1}=0.7731$ (notice the presence of a positive
eigenvalue in the spectrum). Here and in other figures, the blue solid and
red dashed lines represent linearly stable and unstable solutions,
respectively.}
\label{fig:dfo1}
\end{figure}

The top panel in Fig.~\ref{fig:dfo1} shows a typical stable soliton and its
eigenvalue spectrum in the self-defocusing regime with the linear and cubic
gain applied at the HS, $\Gamma _{1}=0.9048$ and $E=-0.16$ (we stress that
the mode is stable in spite of the presence of the \emph{unsaturated}
nonlinear gain). With these parameters, Eqs.~(\ref{eq:sys}) yield $A=0.6958$%
, $\lambda _{1}=0.5615$, $\lambda _{2}=5.2758$, and $k=-0.3795$. When the
linear gain is increased from $\Gamma _{1}=0.9048$ to $\Gamma _{1}=0.9936$,
the stable solution attains its largest amplitude, $A_{\max }=0.8675$, as
shown in the middle panel. With the further increase of the amplitude, an
unstable eigenvalue in the spectrum emerges from the origin into the right
half-plane. The bottom panel depicts such an unstable solution with
amplitude $A=1.074>A_{\max }$, the corresponding linear gain being $\Gamma
_{1}=0.7731$.

\begin{figure}[t]
\begin{center}
\includegraphics[width = 80mm, keepaspectratio]{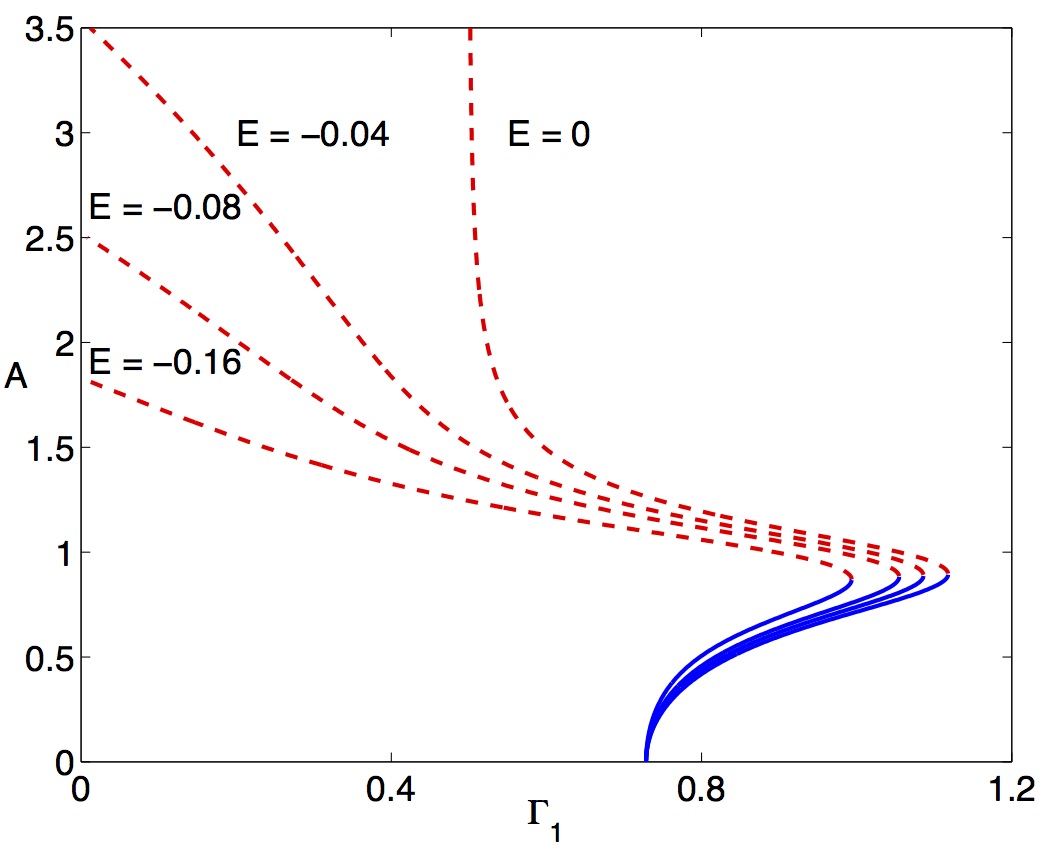}
\end{center}
\caption{(Color online) Amplitude $A$ of the pinned soliton as a function of
linear ($\Gamma _{1}$) and cubic ($E\leq 0$) gain. The other parameters are $%
\protect\gamma =0.5$, $\Gamma _{2}=0.8$, and $B=-1$.}
\label{fig:dfo2}
\end{figure}

Figure~\ref{fig:dfo2} shows amplitude $A$ of the stable (solid) and unstable
(dashed) pinned modes as a function of linear gain $\Gamma _{1}>0$ and cubic
gain $E<0$. In the absence of the cubic dissipation, i.e., at $E=0$, there
exists a stable family of the modes in the region of $0.73\leq \Gamma
_{1}\leq 1.11$, with the amplitude ranging from $A=0.08$ to $A=0.89$.
Outside this stability region, any solution governed by Eq.~(\ref{eq:gl})
either decays to zero, when the linear gain is too small ($\Gamma _{1}<0.73$%
), or blows up to infinity when it is too large ($\Gamma _{1}>1.11$).

Figure~\ref{fig:dfo2} shows that, when the cubic gain increases, the largest
amplitude of the stable soliton and the corresponding linear gain drop, but
only by small amounts. Naturally, the bifurcation of the zero solution $A=0$
into the pinned mode takes place at a particular value $\Gamma _{1}=0.7286$,
regardless of the value of $E$. On the contrary, the unstable branches show
a large variation in amplitude as $E$ varies. At $E=0$, there is a vertical
asymptote of the unstable branch exactly at $\Gamma _{1}=0.5$. This implies
that, if the local linear gain is too weak, i.e., $\Gamma _{1}<0.5$, it
cannot compensate the background loss without the contribution from the
nonlinear gain. Although Fig.~\ref{fig:dfo2} shows the unstable branches
only in the region of $\Gamma _{1}\geq 0$, the curves corresponding to $E<0$
extend to the region of $\Gamma _{1}<0$ (linear loss). These unstable
solitons are supported by the nonlinear gain alone.

\begin{figure}[t]
\begin{center}
\includegraphics[width = 80mm, keepaspectratio]{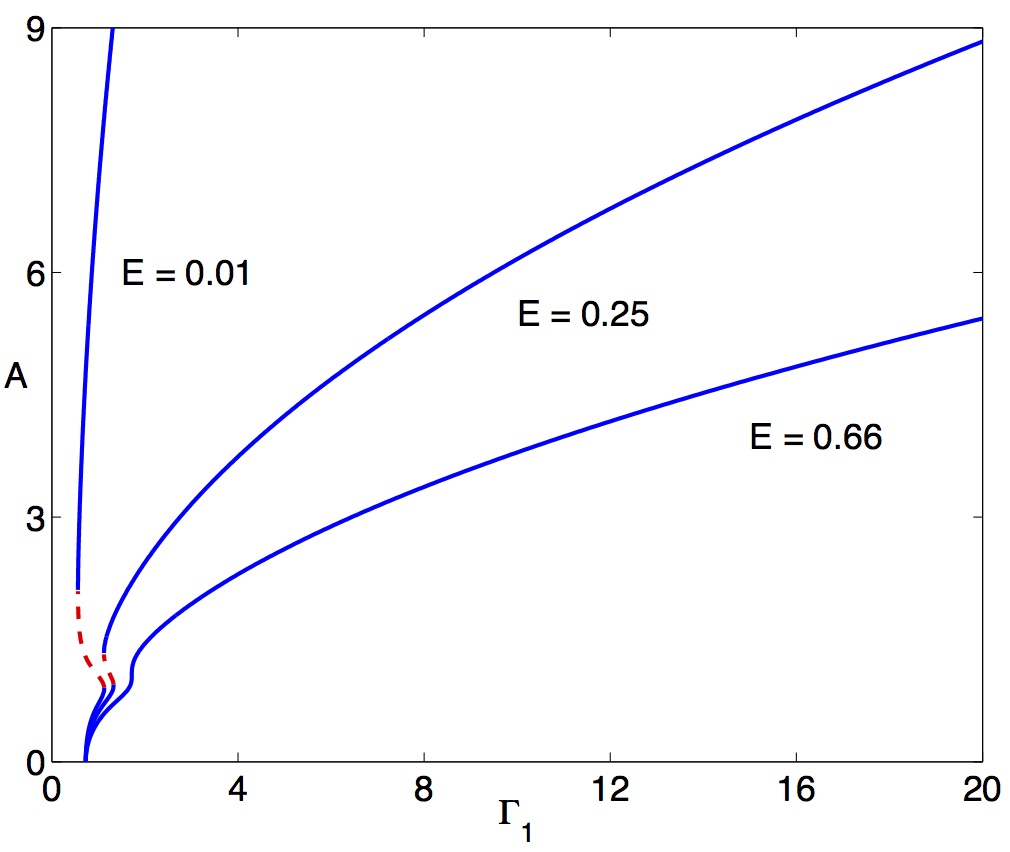}
\end{center}
\caption{(Color online) Solution branches in the case of the cubic loss ($%
E>0 $). The other parameters are same as those in Fig.~\protect\ref{fig:dfo2}.}
\label{fig:dfo3}
\end{figure}

As mentioned in the Introduction, the existence of the stability region for
the pinned modes in the absence of the gain saturation is a remarkable
feature of the system. On the other hand, the stability region is (quite
naturally) much broader in the case of the cubic loss, $E>0$. Figure~\ref%
{fig:dfo3} shows the respective solution branches obtained with the
localized self-defocusing nonlinearity. When the cubic loss is small, e.g., $%
E=0.01$, there are two distinct families of stable modes, representing broad
small-amplitude ($A\leq 0.89$) and narrow large-amplitude ($A\geq 2.11$)
ones. These two stable families are linked by an unstable branch with the
amplitudes in the interval of $0.89<A<2.11$. There is a range of values of
the linear gain, $0.73\leq \Gamma _{1}\leq 1.13$, for which the two stable
branches coexist, thus making the system \emph{bistable}. Figure~\ref%
{fig:dfo4} shows the coexisting stable solutions in the bistability region.
In direct simulations, a localized input evolves into either of these two
stable solutions, depending on the initial amplitude. With the increase of $E
$, the amplitude drops to compensate the growing nonlinear loss, stretching
the solution curves in Fig.~\ref{fig:dfo3} in the horizontal direction.
Simultaneously, the bistability region and the unstable branch shrink.
Eventually, both of them disappear at $E\approx 0.66$.

\begin{figure}[t]
\begin{center}
\includegraphics[width = 80mm, keepaspectratio]{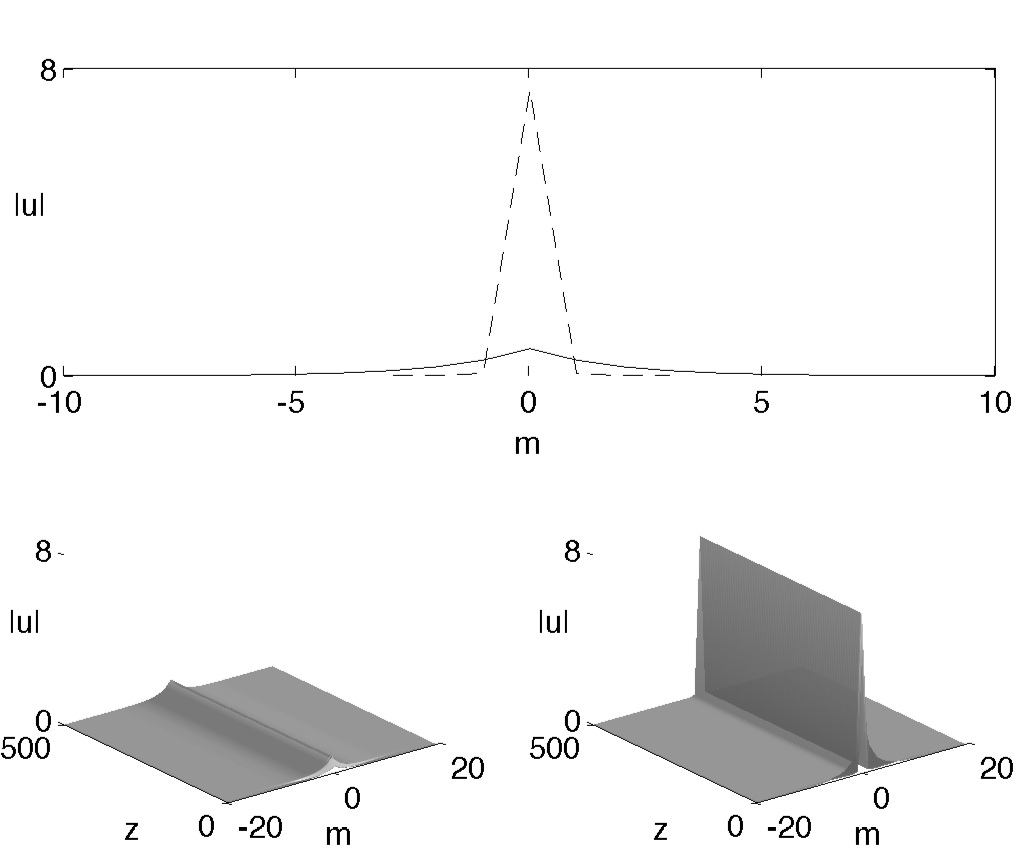}
\end{center}
\caption{The coexistence of stable small- and large-amplitude pinned modes
(top), and the corresponding evolution of perturbed solutions (bottom) at $%
E=0.01$, in the bistability region. Inputs with amplitudes $A=0.3$ and $A=2$
evolve into the small-amplitude and large-amplitude stationary modes,
respectively. The other parameters are $\protect\gamma =0.5$, $\Gamma _{1}=1$%
, $\Gamma _{2}=0.8$, and $B=-1$.}
\label{fig:dfo4}
\end{figure}

\subsection{The self-focusing regime ($B=+1$)}

\begin{figure}[t]
\begin{center}
\includegraphics[width = 80mm, keepaspectratio]{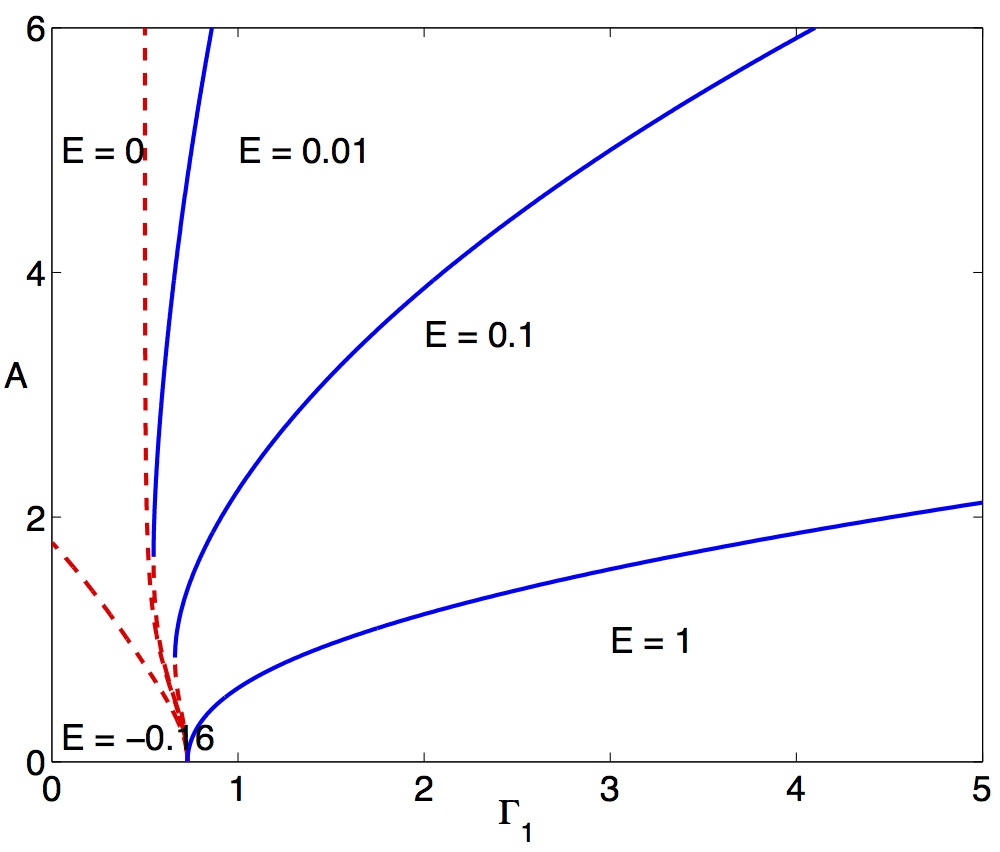}
\end{center}
\caption{(Color online) Amplitude $A$ of the pinned mode as a function of
the linear gain ($\Gamma _{1}$) and cubic loss ($E$) in the model with the
self-focusing nonlinearity. The other parameters are $\protect\gamma =0.5$, $%
\Gamma _{2}=0.8$, and $B=+1$.}
\label{fig:fo1}
\end{figure}

\begin{figure}[t]
\begin{center}
\includegraphics[width = 80mm, keepaspectratio]{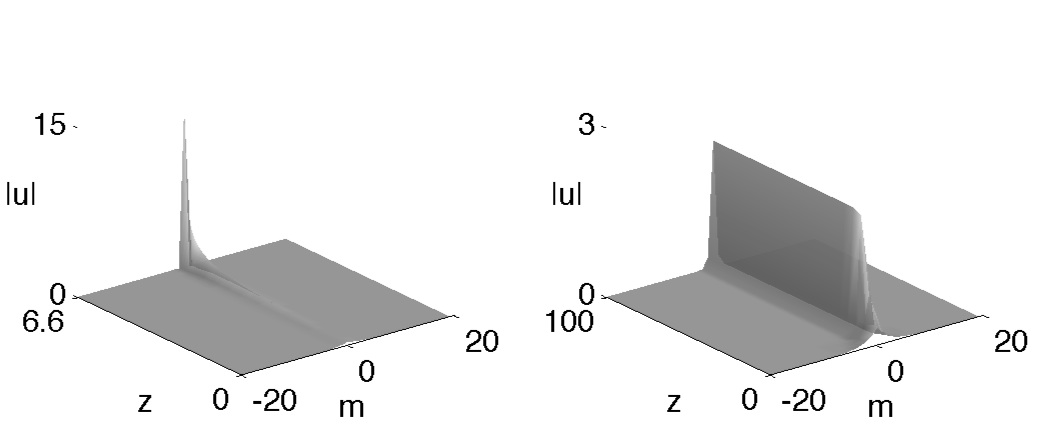}
\end{center}
\caption{The evolution of perturbed solitons at $E=-0.1$ (left) and $E=0.1$
(right). The other parameters are $\protect\gamma =0.5$, $\Gamma _{1}$, $%
\Gamma _{2}=0.8$, and $B=1$. }
\label{fig:fo2}
\end{figure}

For the self-focusing nonlinearity, $B=+1$, the branches of pinned modes~(%
\ref{eq:pw}) are shown in Fig.~\ref{fig:fo1} as functions of $\Gamma _{1}$
and $E$. Under the action of the self-focusing, all the pinned modes are
unstable without the cubic loss, i.e., at $E\leq 0$. The left panel of Fig.~%
\ref{fig:fo2} demonstrates that this instability quickly leads to a blowup
of the dissipative soliton. For the parameters considered here, all the
unstable solutions originate from point $\Gamma _{1}=0.73$. There is a
vertical asymptote at $\Gamma _{1}=0.5$ for the unstable solutions
corresponding to $E=0$. These observations are identical to those made in
the case of the self-defocusing nonlinearity ($B=-1$, see Figs.~\ref%
{fig:dfo2} and~\ref{fig:dfo3}.)

\begin{figure}[t]
\begin{center}
\includegraphics[width = 80mm, keepaspectratio]{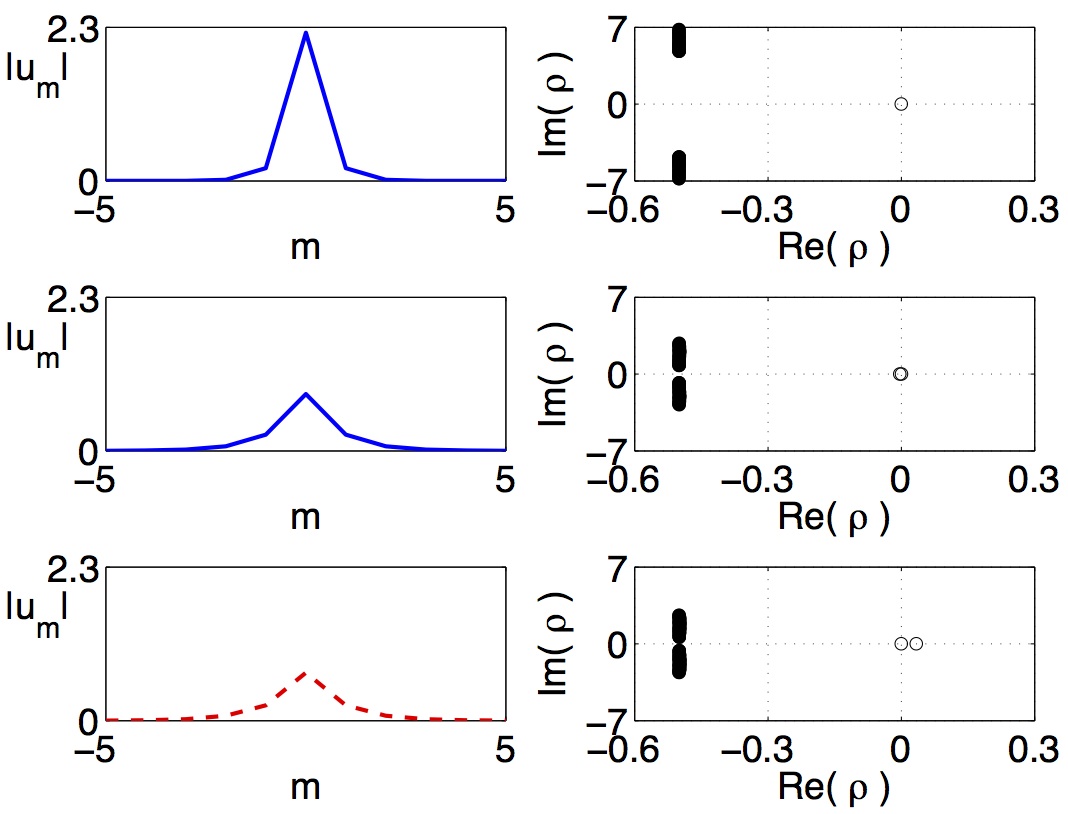}
\end{center}
\caption{(Color online) Examples of stable and unstable pinned solitons
(left) and the corresponding stability spectra (right) in the case of the
self-focusing nonlinearity combined with the cubic loss: $B=1$, $E=0.1$, $%
\protect\gamma =0.5$, and $\Gamma _{2}=0.8$. Top and middle: Stable
solutions found at $\Gamma _{1}=0.9991$ and $\Gamma _{1}=0.6616$,
respectively. Bottom: An unstable solution obtained at $\Gamma _{1}=0.6648$.
In the latter case, there is a positive eigenvalue at $\protect\rho =0.034$.}
\label{fig:fo3}
\end{figure}

As shown in the right panel of Fig.~\ref{fig:fo2}, the dynamical blowup is
naturally prevented by the cubic loss ($E>0$). Figure~\ref{fig:fo1} shows
the solution branches for this case too. When linear gain $\Gamma _{1}$
falls below a certain threshold, the modes do not exist, as the background
loss cannot be compensated. In this case, all initial condition decay to
zero. Once $\Gamma _{1}$ exceeds the threshold, the system supports the
localized modes, which remain stable even at very large values of $\Gamma
_{1}$. Figure~\ref{fig:fo3} shows some representative examples. For
instance, with linear gain $\Gamma _{1}=0.9991$, the system supports a
stable pinned mode of amplitude $A=2.217$ (the top panel). The stable
solution with the smallest amplitude ($A=0.8521$) is found at $\Gamma
_{1}=0.6616$ (the middle panel). The solutions with amplitudes $A<0.8521$
are unstable---for instance, the one shown in the bottom panel of Fig.~\ref%
{fig:fo3}

\subsection{Zero Kerr Nonlinearity ($B = 0$)}

\begin{figure}[t]
\begin{center}
\includegraphics[width = 80mm, keepaspectratio]{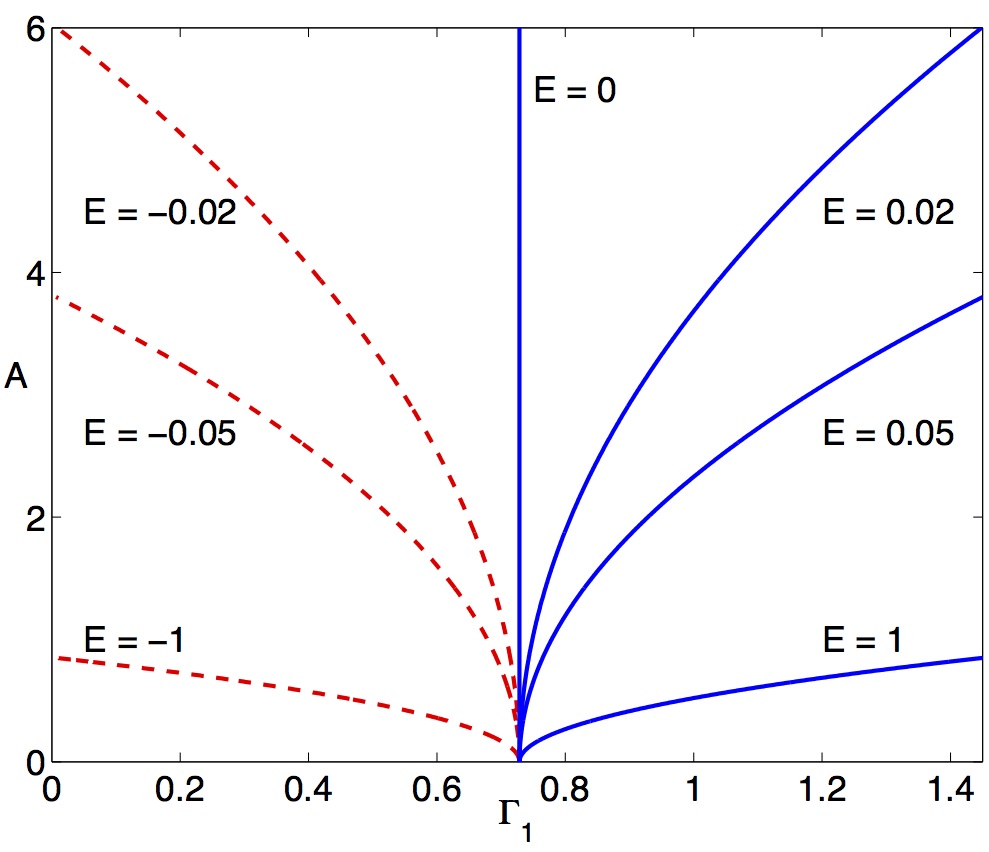}
\end{center}
\caption{(Color online) Solution branches as a function of the cubic gain or
loss $E$ in the case of $B=0$. The other parameters are the same as those in Fig.~%
\protect\ref{fig:fo2}.}
\label{fig:fo4}
\end{figure}

\begin{figure}[t]
\begin{center}
\includegraphics[width = 80mm, keepaspectratio]{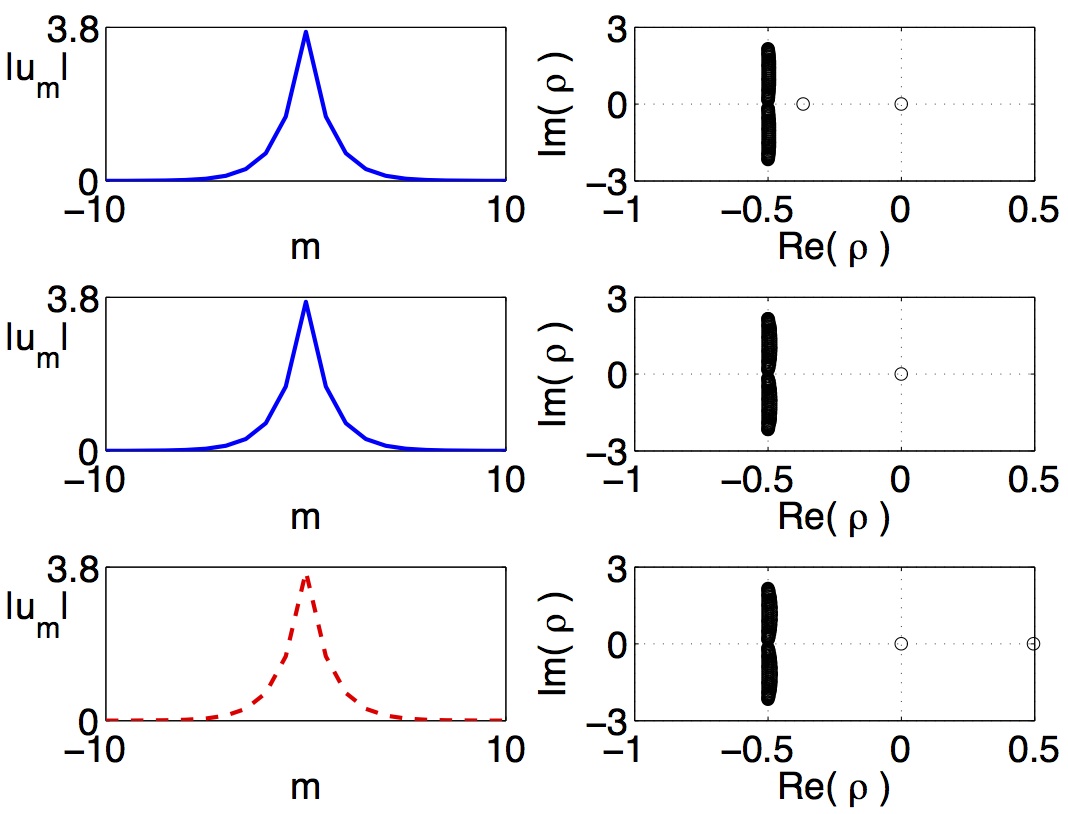}
\end{center}
\caption{(Color online) Examples of stable and unstable pinned modes (left)
and the corresponding stability spectra (right) in the absence of the Kerr
nonlinearity ($B=0$). All these solutions have the same amplitudes, $A=3.687$%
. The other parameters are $\protect\gamma =0.5$, and $\Gamma _{2}=0.8$.
Top: A stable solution found at $E=0.02$ with $\Gamma _{1}=1.0004$. Middle:
A stable solution found in the linear system, with $E=0$ and $\Gamma
_{1}=0.73$ (the amplitude is arbitrary in this case). Bottom: An unstable
solution obtained at $E=-0.02$ with $\Gamma _{1}=0.4568$. In the latter
case, there is a positive eigenvalue, $\protect\rho =0.4955$.}
\label{fig:fo5}
\end{figure}

We have also studied the pinned modes in the case of $B=0$, when the
nonlinearity at the HS is represented solely by the cubic gain or loss, $%
E\neq 0$. Figure~\ref{fig:fo4} shows the respective solution branches
corresponding to different values of $E$. The linear stability analysis
shows that all these solutions are unstable in the presence of the cubic
gain ($E<0$), while the solutions corresponding to the nonlinear loss, $E>0$%
, are always stable. An interesting feature found here is that the stable
and unstable branches in the parameter space are symmetric about the
solution for $E=0$. This particular branch has an arbitrary amplitude, as
there is no nonlinearity when both $B$ and $E$ vanish. All the solutions
belonging to this branch correspond to linear gain $\Gamma _{1}=0.73$, which
is again the critical value at which the zero background bifurcates into the
pinned mode (see Figs.~\ref{fig:dfo2}, \ref{fig:dfo3} and~\ref{fig:fo1}).
Several stable and unstable modes and their linear spectra are shown for the
case of $B=0$ and different values of $E$ in Fig.~\ref{fig:fo5}.

\subsection{Stability and instability of the pinned modes with respect to
finite perturbations}

\begin{figure}[t]
\begin{center}
\includegraphics[width = 80mm, keepaspectratio]{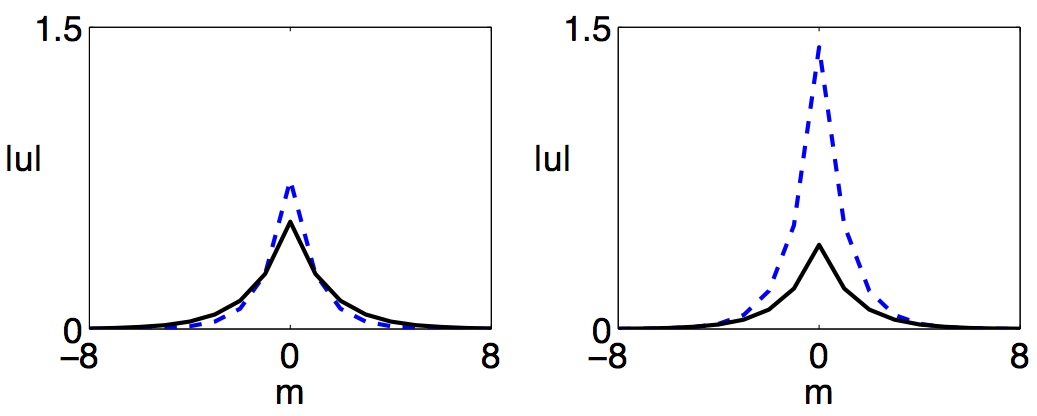}
\end{center}
\caption{(Color online) The blue dashed curves show the largest
initial disturbance, corresponding to the sudden ``stretch" of the
soliton, after the application of which the pinned dissipative
soliton still relaxes back into the original form. A yet stronger
stretch quickly initiates a blowup. The left and right panels
pertain to $E=-0.2$ and $0$, respectively. The other parameters
being $B=-1$, and $\Gamma_1 = \Gamma_2 = 0.8$.} \label{add1}
\end{figure}

It has been shown above that, while the pinned modes may be stable
against infinitesimal perturbations under the combined action of the
self-defocusing nonlinearity ($B=-1$) and unsaturated nonlinear gain
($E\leq 0$), the respective stability area is much smaller than that in
the case of the nonlinear loss ($E>0$), see Figs. \ref{fig:dfo2} and
\ref{fig:dfo3}. The apparent fragility of the dissipative solitons
in the case of $E\leq 0$, and their robustness at $E>0$ suggest an investigation to
check the stability of these two types of the pinned modes against
finite-amplitude perturbations, which we have carried out by means
of systematic simulations of the evolution of the modes perturbed by reasonably large initial disturbances where linearization will not be an adequate approximation initial disturbances. The conclusion is that the ``robust"
solitons, found at $E>0$, are completely stable against arbitrary,
finite amplitude perturbations. On the other hand, one can always destroy
the ``fragile" modes, which are stable against infinitesimal
perturbations at $E\leq 0$, by applying perturbations of a
sufficiently large amplitude. If, in particular, the finite
disturbance is applied by suddenly making its amplitude larger than
it is in the stationary solution (``stretching"), the soliton will
blow up if the stretching factor exceeds a particular critical
value. The corresponding instability borders for the solitons with
$E=-0.2$ and $E=0$ are displayed in Fig. \ref{add1}.

\begin{figure}[t]
\begin{center}
\includegraphics[width = 80mm, keepaspectratio]{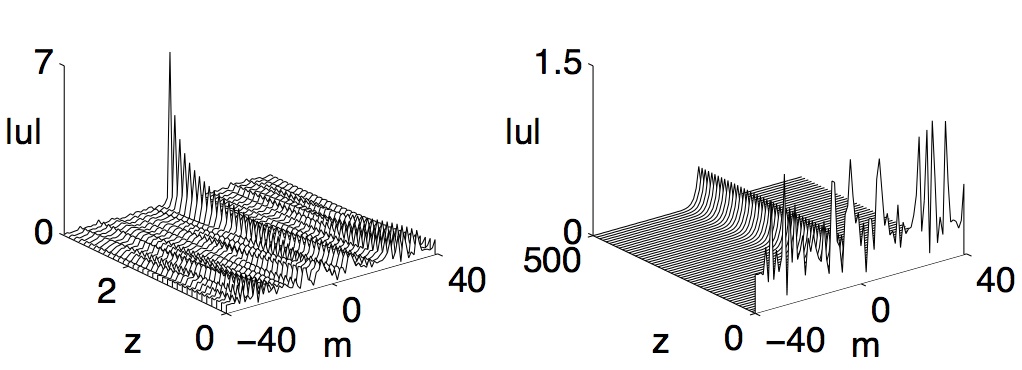}
\end{center}
\caption{ The left panel: The evolution of a strongly
perturbed ``fragile" pinned mode, at $E=-0.2$. The right panel: the
same, but for $E=+0.2$. Other parameters are as in Fig. \protect
\ref{add1}. The other parameters are same as those used in Fig.~\ref{add1}.} \label{add2}
\end{figure}

In addition, the blowup of the linearly stable but ``fragile" pinned
mode, caused by arbitrary finite amplitude perturbations, and,
simultaneously, the full stability of the ``robust" mode against
still stronger perturbations, are illustrated by their evolution
histories presented in Fig. \ref{add2}. It may be concluded that the
entire space of the initial conditions is the attraction basin of
the latter mode, while for the one supported by the unsaturated
cubic gain the attraction basin is quite narrow.

\section{Conclusions}

\label{sec:con}

We have introduced the discrete dynamical system based on the linear
lossy lattice into which a single nonlinear site with the linear
gain (HS, ``hot spot") is embedded. The system can be readily
implemented in the form of an array of optical or plasmonic
waveguides, admitting selective excitation of individual cores, by
the local application of the pump to the uniformly doped cores.
Solutions for solitons pinned to the central site were found in the
implicit analytical form, and their stability against infinitesimal and finite perturbations was investigated numerically.
Stability regions for the solitons have been identified in the
parameter plane of the most essential control parameters of the
system, \emph{viz}., the linear gain $\Gamma _{1}$ and cubic
dissipation $E$. A nontrivial finding is a (rather small) stability
area for the solitons supported by the combination of the local
nonlinear \emph{unsaturated} gain and self-defocusing cubic
nonlinearity. On the other hand, the combination of the cubic loss
and self-defocusing nonlinearity gives rise to the bistability of
the pinned solitons. In the former case, the collapse of the
linearly stable soliton is caused by finite amplitude perturbations. These
features may be promising for potential applications, and call for
an experimental implementation.

The work may be naturally extended in different directions. In particular,
it will be interesting to investigate localized modes pinned by pairs of hot
spots (cf.the analysis of
a discrete counterpart of the continuous model of Ref. \cite{spotsExact2}). A challenging possibility is to develop the analysis for
modes pinned to the hot spot embedded into a two-dimensional linear lossy
lattice. In that case, an analytical solution is not available even for the
linear lattice, hence the entire analysis should be done in a numerical form.

\begin{acknowledgements}
B. A. Malomed and E. Ding acknowledge the hospitality of the Department of
Mechanical Engineering at the University of Hong Kong.
Partial financial support was provided by the Hong Kong Research Grants Council.
\end{acknowledgements}

\end{document}